\documentclass[a4paper,11pt]{article}
\pdfoutput=1 

\usepackage{jheppub} 

\usepackage[T1]{fontenc} 

\newcommand{\Scp}{\mathcal{S}_{\eta'K^0}}
\newcommand{\Acp}{\mathcal{A}_{\eta'K^0}}
\newcommand{\seff}{\sin 2\phi^{\rm eff}_1}
\newcommand{\gmm}{~{\rm GeV}/c^2}

\newcommand{\T}{\rule{0pt}{2.6ex}}       
\newcommand{\B}{\rule[-1.2ex]{0pt}{0pt}} 

\preprint{\vbox{ \hbox{   }
    \hbox{Belle Preprint 2014-10}
    \hbox{KEK Preprint 2014-17}
    \hbox{October 25, 2014}
}}

\title{\boldmath Measurement of Time-Dependent $CP$ Violation in $B^0\to \eta'K^0$ Decays}

\collaborationImg{\includegraphics[width=.1\textwidth]{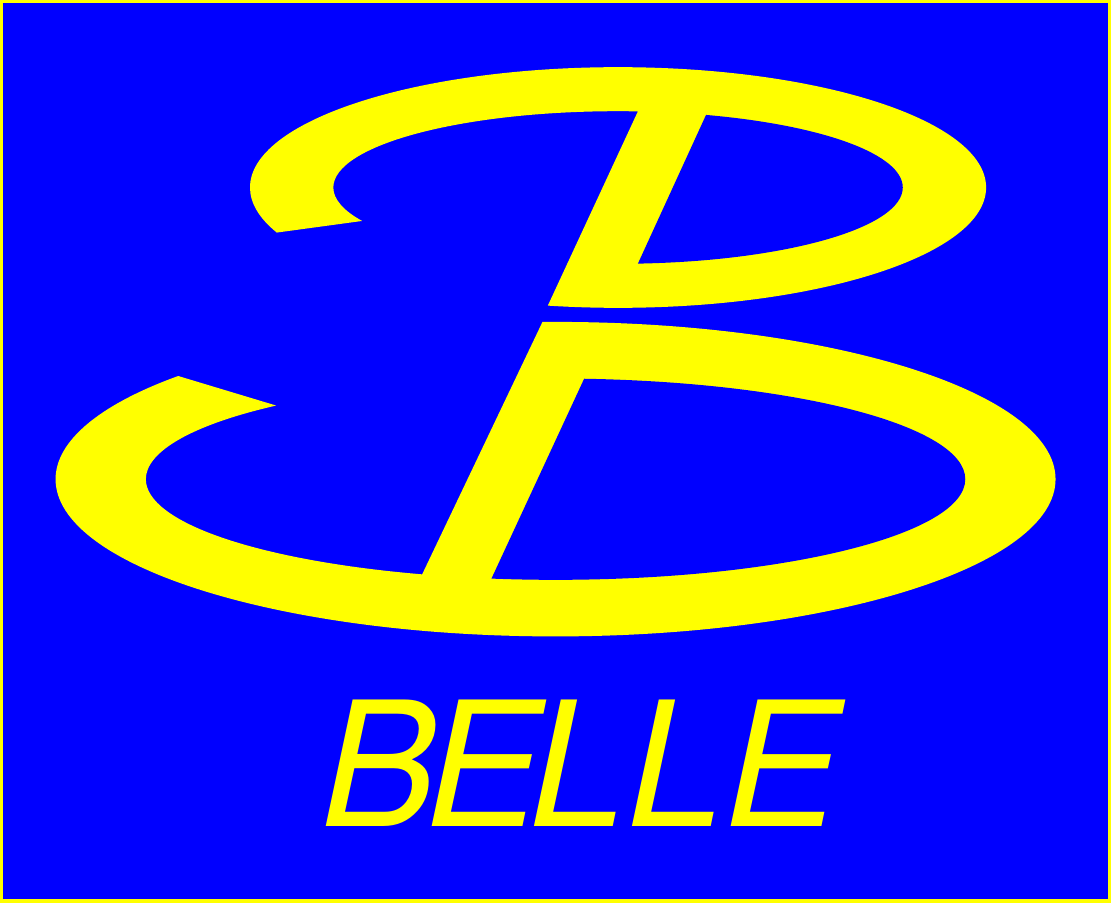}}
\affiliation[1]{University of the Basque Country UPV/EHU, 48080 Bilbao}
\affiliation[2]{Beihang University, Beijing 100191}
\affiliation[3]{University of Bonn, 53115 Bonn}
\affiliation[4]{Budker Institute of Nuclear Physics SB RAS and Novosibirsk State University, Novosibirsk 630090}
\affiliation[5]{Faculty of Mathematics and Physics, Charles University, 121 16 Prague}
\affiliation[6]{Chonnam National University, Kwangju 660-701}
\affiliation[7]{University of Cincinnati, Cincinnati, Ohio 45221}
\affiliation[8]{Deutsches Elektronen--Synchrotron, 22607 Hamburg}
\affiliation[9]{Justus-Liebig-Universit\"at Gie\ss{}en, 35392 Gie\ss{}en}
\affiliation[10]{Gifu University, Gifu 501-1193}
\affiliation[11]{The Graduate University for Advanced Studies, Hayama 240-0193}
\affiliation[12]{Hanyang University, Seoul 133-791}
\affiliation[13]{University of Hawaii, Honolulu, Hawaii 96822}
\affiliation[14]{High Energy Accelerator Research Organization (KEK), Tsukuba 305-0801}
\affiliation[15]{IKERBASQUE, Basque Foundation for Science, 48011 Bilbao}
\affiliation[16]{Indian Institute of Technology Bhubaneswar, Satya Nagar 751007}
\affiliation[17]{Indian Institute of Technology Guwahati, Assam 781039}
\affiliation[18]{Indian Institute of Technology Madras, Chennai 600036}
\affiliation[19]{Indiana University, Bloomington, Indiana 47408}
\affiliation[20]{Institute of High Energy Physics, Chinese Academy of Sciences, Beijing 100049}
\affiliation[21]{Institute of High Energy Physics, Vienna 1050}
\affiliation[22]{Institute for High Energy Physics, Protvino 142281}
\affiliation[23]{INFN - Sezione di Torino, 10125 Torino}
\affiliation[24]{Institute for Theoretical and Experimental Physics, Moscow 117218}
\affiliation[25]{J. Stefan Institute, 1000 Ljubljana}
\affiliation[26]{Kanagawa University, Yokohama 221-8686}
\affiliation[27]{Institut f\"ur Experimentelle Kernphysik, Karlsruher Institut f\"ur Technologie, 76131 Karlsruhe}
\affiliation[28]{Kavli Institute for the Physics and Mathematics of the Universe (WPI), University of Tokyo, Kashiwa 277-8583}
\affiliation[29]{Department of Physics, Faculty of Science, King Abdulaziz University, Jeddah 21589}
\affiliation[30]{Korea Institute of Science and Technology Information, Daejeon 305-806}
\affiliation[31]{Korea University, Seoul 136-713}
\affiliation[32]{Kyungpook National University, Daegu 702-701}
\affiliation[33]{\'Ecole Polytechnique F\'ed\'erale de Lausanne (EPFL), Lausanne 1015}
\affiliation[34]{Faculty of Mathematics and Physics, University of Ljubljana, 1000 Ljubljana}
\affiliation[35]{Luther College, Decorah, Iowa 52101}
\affiliation[36]{University of Maribor, 2000 Maribor}
\affiliation[37]{Max-Planck-Institut f\"ur Physik, 80805 M\"unchen}
\affiliation[38]{School of Physics, University of Melbourne, Victoria 3010}
\affiliation[39]{Moscow Physical Engineering Institute, Moscow 115409}
\affiliation[40]{Moscow Institute of Physics and Technology, Moscow Region 141700}
\affiliation[41]{Graduate School of Science, Nagoya University, Nagoya 464-8602}
\affiliation[42]{Kobayashi-Maskawa Institute, Nagoya University, Nagoya 464-8602}
\affiliation[43]{Nara Women's University, Nara 630-8506}
\affiliation[44]{National Central University, Chung-li 32054}
\affiliation[45]{National United University, Miao Li 36003}
\affiliation[46]{Department of Physics, National Taiwan University, Taipei 10617}
\affiliation[47]{H. Niewodniczanski Institute of Nuclear Physics, Krakow 31-342}
\affiliation[48]{Niigata University, Niigata 950-2181}
\affiliation[49]{Osaka City University, Osaka 558-8585}
\affiliation[50]{Pacific Northwest National Laboratory, Richland, Washington 99352}
\affiliation[51]{Peking University, Beijing 100871}
\affiliation[52]{University of Pittsburgh, Pittsburgh, Pennsylvania 15260}
\affiliation[53]{University of Science and Technology of China, Hefei 230026}
\affiliation[54]{Seoul National University, Seoul 151-742}
\affiliation[55]{Soongsil University, Seoul 156-743}
\affiliation[56]{Sungkyunkwan University, Suwon 440-746}
\affiliation[57]{School of Physics, University of Sydney, NSW 2006}
\affiliation[58]{Department of Physics, Faculty of Science, University of Tabuk, Tabuk 71451}
\affiliation[59]{Tata Institute of Fundamental Research, Mumbai 400005}
\affiliation[60]{Excellence Cluster Universe, Technische Universit\"at M\"unchen, 85748 Garching}
\affiliation[61]{Toho University, Funabashi 274-8510}
\affiliation[62]{Tohoku University, Sendai 980-8578}
\affiliation[63]{Department of Physics, University of Tokyo, Tokyo 113-0033}
\affiliation[64]{Tokyo Institute of Technology, Tokyo 152-8550}
\affiliation[65]{Tokyo Metropolitan University, Tokyo 192-0397}
\affiliation[66]{University of Torino, 10124 Torino}
\affiliation[67]{CNP, Virginia Polytechnic Institute and State University, Blacksburg, Virginia 24061}
\affiliation[68]{Wayne State University, Detroit, Michigan 48202}
\affiliation[69]{Yamagata University, Yamagata 990-8560}
\affiliation[70]{Yonsei University, Seoul 120-749}
\author[14]{L.~\v{S}antelj,}
  \author[48]{Y.~Yusa,}
  \author[58]{A.~Abdesselam,}
  \author[14]{I.~Adachi,}
  \author[63]{H.~Aihara,}
  \author[58]{S.~Al~Said,}
  \author[50]{D.~M.~Asner,}
  \author[4]{V.~Aulchenko,}
  \author[24]{T.~Aushev,}
  \author[58]{R.~Ayad,}
% ...
\collaboration{The Belle Collaboration}

\emailAdd{luka.santelj@kek.jp}
\emailAdd{yusa@hep.sc.niigata-u.ac.jp}

\abstract{
We present a measurement of the time-dependent $CP$ violation parameters in $B^0\to\eta'K^0$ decays. The measurement is based on the full data sample containing $772\times 10^6$ $B\bar{B}$ pairs collected at the $\Upsilon(4S)$ resonance using the Belle detector at the KEKB asymmetric-energy $e^+e^-$ collider. The measured values of the mixing-induced and direct $CP$ violation parameters are:
\begin{align} 
  \sin 2\phi^{\rm eff}_1 &= +0.68\pm 0.07 \pm 0.03, \nonumber \\
\mathcal{A}_{\eta'K^0} &= +0.03\pm 0.05\pm 0.04, \nonumber 
\end{align}
where the first uncertainty is statistical and the second is systematic. The values obtained are the most accurate to date. Furthermore, these results are consistent with our previous measurements and with the world-average value of $\sin 2\phi_1$ measured in $B^0\to J/\psi K^0$ decays.}

\begin{document} 
\maketitle
\flushbottom

\section{Introduction}

{\it CP} violation in the quark sector is described within the Standard Model (SM) by a single irreducible complex phase in the Cabibbo-Kobayashi-Maskawa (CKM) quark mixing matrix \cite{km}. Unitarity of the CKM matrix gives rise to six so-called unitarity triangles in the complex plane. One is related to transition amplitudes involving the $b$ quark and is characterized by three large angles $\phi_1$, $\phi_2$ and $\phi_3$. In the past decade, determination of the value of $\sin 2\phi_1$, mainly by the measurements of time-dependent {\it CP} asymmetries in $B^0$ decays that are dominated by the $b\to c\bar{c}s$ quark transition \cite{belle_cpv,babar_cpv}, has provided an important test and confirmation of the Kobayashi-Maskawa (KM) mechanism. Despite the great success of the KM mechanism, which in principle gives rise to a matter-antimatter asymmetry in the Universe, new sources of {\it CP} violation are required to account for the magnitude of the observed asymmetry \cite{bgn}. Promising places to search for additional {\it CP} violating effects are $B^0$ meson decays dominated by the $b\to s\bar{q}q$ quark transition, which, in the SM, proceeds through a single loop (penguin) diagram, and is therefore sensitive to possible new heavy particle contributions in the loop \cite{loopcont1,loopcont2,loopcont3}. The decay $B^0\to\eta'K^0$ studied here belongs to this category. 

The KM mechanism predicts a {\it CP} asymmetry in the time-dependent decay rates for $B^0$ and $\bar{B}^0$ to {\it CP} eigenstates \cite{carter,bigi} --- in our case, $\eta'K^0_S$ and $\eta'K^0_L$. The $B$ factories operated at the $\Upsilon(4S)$ resonance, which decays almost exclusively into correlated $B\bar{B}$ pairs. In the decay chain $\Upsilon(4S)\to B^0\bar{B}^0\to f_{\rm rec}~ f_{\rm tag}$, where the reconstructed $B$ meson decays into $f_{\rm rec}=\eta'K^0$ at time $t_{\rm rec}$ and the tagging $B$ meson decays into $f_{\rm tag}$ at time $t_{\rm tag}$, the distribution of the decay time difference $\Delta t = t_{\rm rec} - t_{\rm tag}$ is given by
\begin{equation}
\mathcal{P}(\Delta t,q) = \frac{e^{-|\Delta t|/\tau_{B^0}}}{4\tau_{B^0}} \left( 1 + q \cdot \left[ \Scp\sin (\Delta m_d \Delta t) + \Acp \cos (\Delta m_d \Delta t)  \right] \right).
\label{eq:sig_pdf}
\end{equation}
Here $\tau_{B^0}$ is the $B^0$ lifetime, $\Delta m_{d}$ is the mass difference between the two neutral $B^0$ mass eigenstates, $q = +1~(-1)$ when the tagging $B$ meson is a $B^0$ ($\bar{B}^0$), and $\Scp$ and $\Acp$ are the ${\it CP}$ violation parameters. Assuming the $b\to s\bar{q}q$ penguin amplitude dominates the $B^0\to\eta'K^0$ decay, the SM expectation is $\Scp=-\xi_f\sin 2\phi_1$ and $\Acp=0$, where $\xi_f$ is the {\it CP} eigenvalue of the final state, $-1~ (+1)$ for the $\eta'K^0_S$ ($\eta'K^0_L$) final state. 
Here, we denote the mixing-induced $CP$ violation parameter as $\sin 2\phi^{\rm eff}_1 = -\xi_f\mathcal{S}_{\eta'K^0}$. Note that the contributions from the CKM-suppressed amplitudes and the color-suppressed $b\to u$ tree amplitude to the decay may result in $\sin 2\phi^{\rm eff}_1$ deviating from $\sin 2\phi_1$ as determined by measurements of $b\to c\bar{c}s$ decays and also induce non-zero $\Acp$ even in the SM. To estimate the possible size of the deviation $\Delta \Scp = \sin 2\phi^{\rm eff}_1 -\sin 2\phi_1$ within the SM, several theoretical approaches are used. For example, the $SU(3)_F$ approach limits $\Delta \Scp$ to the range $[-0.05,0.09]$ \cite{su31}, while QCD factorization constrains it to $[-0.03,0.03]$ \cite{qcdf1,qcdf2,qcdf3}. Other calculations can be found in Refs. \cite{smcalc1,smcalc2,smcalc3}; these produce values close to those quoted above. Observing values of $\Delta \Scp$ significantly larger than these predictions  would be a sign of new physics contributions. In previous measurements of $\seff$ and $\Acp$ by the Belle \cite{etap_belle} and the BaBar \cite{etap_babar} Collaborations, no significant deviations from the SM predicted values were observed. However, their rather large statistical uncertainties ($\sim 0.1$) motivate more precise measurements.      

In this paper, we present an updated measurement of the parameters $\seff$ and $\Acp$ using the full Belle data set with $772\times 10^6$ $B\bar{B}$ pairs; this supersedes our previous analysis that used $534\times 10^6$ $B\bar{B}$ pairs \cite{etap_belle}. The larger data sample and improved track reconstruction and event selection methods result in a number of reconstructed signal events in this measurement that is almost twice as large as the previous one, while maintaining comparable sample purity. This paper is organized as follows: in section \ref{sec:belle}, we briefly describe the Belle detector and the data sample. The event reconstruction (including vertex and flavor reconstruction) and the event selection criteria are described in section \ref{sec:reco}. In section \ref{sec:result}, we present the measurement results, their systematic uncertainties, and the method validation tests. We conclude with a summary in section \ref{sec:summary}.

\section{The Belle detector and data sample}
\label{sec:belle}
The measurement presented here is based on a data sample containing $772\times 10^6$ $B\bar{B}$ pairs collected at the $\Upsilon(4S)$ resonance with the Belle detector at the KEKB asymmetric-energy $e^+e^-$ (3.5 GeV on 8.0 GeV) collider \cite{kekb,kekb1}. At KEKB, $B\bar{B}$ pairs are  produced with a Lorentz boost of $\beta\gamma=0.425$ nearly along the $+z$ direction, which is opposite the positron beam direction.

The Belle detector is a large-solid-angle magnetic spectrometer consisting of a silicon vertex detector (SVD), a 50-layer central drift chamber (CDC), an array of aerogel threshold Cherenkov counters (ACC), a barrel-like arrangement of time-of-flight scintillation counters (TOF), and an electromagnetic calorimeter (ECL) with CsI(Tl) crystals located inside a superconducting solenoid coil that provides a 1.5 T magnetic field. An iron flux-return yoke (KLM) located outside of the coil is instrumented to detect $K^0_L$ mesons and to identify muons. The detector is described in detail elsewhere \cite{belle,belle1}. The data sample used was collected with two different inner detector configurations. The first $152\times 10^6$ of $B\bar{B}$ pairs were collected with a 2.0-cm-radius beampipe and a three-layer silicon vertex detector (SVD1), while the remaining $620\times 10^6$ $B\bar{B}$ pairs were collected with a 1.5-cm-radius beampipe, a four-layer silicon vertex detector (SVD2), and an additional small-cell inner drift chamber. The latter data sample has been reprocessed using a new charged track reconstruction algorithm that significantly increased the event reconstruction efficiency ($\sim 15\%$ higher than that of our previous measurement due to reprocessing alone). 

A large sample of Monte Carlo (MC) simulated events is used to study the distributions of background from $B\bar{B}$ events and to determine the signal event distributions needed to obtain the signal yield. We use the EVTGEN \cite{evtgen} event generator, the output of which is fed into a detailed detector simulation based on the GEANT3 \cite{geant} platform.  

\section{Event Reconstruction and Selection}
\label{sec:reco}
\subsection{Signal Reconstruction}
We reconstruct $B^0$ meson candidates from an $\eta'$ candidate and a $K^0$ candidate, where the latter is reconstructed either as a $K^0_S$ or a $K^0_L$. Charged tracks that are used for $\eta'$ reconstruction, reconstructed within the CDC and SVD, are required to originate from the interaction point (IP). To distinguish charged kaons from pions, we use a kaon (pion) likelihood $\mathcal{L}_{K (\pi)}$ which is formed based on the information from the TOF, ACC, and $dE/dx$ measurements in the CDC. We form a likelihood ratio $\mathcal{R}_{\pi/K} = \mathcal{L}_{\pi} /(\mathcal{L}_{\pi} + \mathcal{L}_{K})$; candidates with $\mathcal{R}_{\pi/K}<0.9$ are classified as pions. With this requirement, we retain $99\%$ of the pion tracks and reject $90\%$ of the kaon tracks. Photons are identified as isolated ECL clusters without associated charged tracks. In the next two subsections, we describe in more detail the $B$ candidate reconstruction with $K^0_S$ and $K^0_L$ in the final state, respectively. All event selection criteria are optimized to minimize the statistical uncertainty on the extracted values of the $CP$ violation parameters.

\subsection*{A. $B^0\to\eta'K^0_S$}   
The $B^0\to\eta'K^0_S$ decay is reconstructed using $K^0_S$ decays to $\pi^+\pi^-$ or $\pi^0\pi^0$. To reconstruct $K^0_S\to\pi^+\pi^-$ decays, we use pairs of oppositely charged pion tracks that have an invariant mass within $0.020 \gmm$ ($3\sigma$) of the $K^0_S$ mass. To further suppress false $K^0_S$ candidates, we use a neural-network-based selection that mainly utilizes the measured flight length of the $K^0_S$ candidate. The variables with the highest signal/background separation power are the distance between the $K^0_S$ decay vertex and the IP in the $x-y$ plane, and the angle between the $K^0_S$ candidate's flight direction and momentum. To select $K^0_S\to \pi^0\pi^0$ decays, we reconstruct $\pi^0$ candidates from pairs of photons with $E_\gamma > 0.05$ GeV, where $E_\gamma$ is the photon energy measured with the ECL. Assuming that the photons originate from the IP, photon pairs with an invariant mass between $0.08 \gmm $ and $0.15 \gmm $ and momentum above $0.1~{\rm GeV}/c$ are treated as $\pi^0$ candidates. To obtain $K^0_S$ candidates from the $\pi^0$ pairs, we perform a kinematic fit with the following constraints: the invariant masses of the two photon pairs are set to the $\pi^0$ nominal mass, all four photons are constrained to arise from a common vertex (which can be displaced from the IP), and the resulting $K^0_S$ is constrained to originate from the reconstructed $B^0$ decay vertex.


For the reconstruction of the $\eta'$ candidate, three decay chains are used: $\eta'\to\rho^0\gamma$ with $\rho^0\to\pi^+\pi^-$, $\eta'\to \eta \pi^+\pi^-$ with $\eta\to\gamma\gamma$, and $\eta'\to \eta \pi^+\pi^-$ with $\eta\to\pi^+\pi^-\pi^0$. In the following, we denote these modes as $\eta'\to \rho^0\gamma$, $\eta'\to\eta(\gamma\gamma)\pi^+\pi^-$, and $\eta'\to \eta(3\pi)\pi^+\pi^-$, respectively. The last mode is not used in combination with $K^0_S$ candidates reconstructed from $\pi^0$ pairs, due to its low signal yield and large background. For the $\eta'\to\rho^0\gamma$ mode, candidate $\rho^0$ mesons are reconstructed from pairs of common-vertex-constrained $\pi^+\pi^-$ tracks, with an invariant mass between $0.50 \gmm$ and $0.95 \gmm$. They are combined with a photon of energy above $0.1 \mbox{ GeV}$, and $\pi^+\pi^-\gamma$ combinations with an invariant mass between $0.932 \gmm$ and $0.975 \gmm$ ($1.7\sigma$) are selected as $\eta'$ candidates. For the modes with an intermediate $\eta$, candidate $\eta$ mesons are formed from photon pairs with an invariant mass between $0.5 \gmm$ and $0.58 \gmm$ ($1.6\sigma$), or $\pi^+\pi^-\pi^0$ combinations with an invariant mass between $0.535 \gmm$ and $0.558 \gmm$ ($2\sigma$). 
Additionally, a kinematic fit is performed with constraints on the $\eta$ mass and its decay vertex as ascertained from the fitted vertex of the $\pi^+\pi^-$ tracks from the $\eta'$ in order to improve the energy resolution. Combinations of reconstructed $\eta$ candidates and $\pi^+\pi^-$ tracks with an invariant mass between $0.942 \gmm$ and $0.970 \gmm$ ($2.5\sigma$) for the $\eta'\to\eta(\gamma\gamma)\pi^+\pi^-$ mode, or between $0.945 \gmm$ and $0.970 \gmm$ ($2.5\sigma$) for the $\eta'\to\eta(3\pi)\pi^+\pi^-$ mode, are selected as $\eta'$ candidates. A kinematic fit with the $\eta'$ mass constraint is performed prior to combining the $\eta'$ and $K^0_S$ to form the $B^0$ candidate.  

Reconstructed $B^0\to\eta'K^0_S$ candidates are identified using the beam-energy-constrained mass $M_{\rm bc} = \sqrt{(E^{*}_{\rm beam})^2 -(p^{*}_B)^2 }$ and the energy difference $\Delta E = E^{*}_{B} - E^{*}_{\rm beam}$, where $E^{*}_{\rm beam}$ is the beam energy in the center-of-mass system (cms) and $E^{*}_B$ and $p^{*}_B$ are the measured cms energy and momentum, respectively, of the reconstructed $B$ candidate. The $M_{\rm bc}$ resolution is about $2.5 ~\rm{MeV}/c^2$, common to all decay modes (since it is dominated by the spread of $E^{*}_{\rm beam}$), while the $\Delta E$ resolution varies from $\sim 20~\rm{MeV}$ for modes with $K^0_S\to\pi^+\pi^-$, to $\sim 50~\rm{MeV}$ for modes with $K^0_S\to\pi^0\pi^0$. 


\subsection*{B. $B^0\to\eta'K^0_L$}

For $B^0 \to \eta' K^0_L$ decays, $K_L^0$ candidates are reconstructed from hit clusters in the KLM and neutral clusters in the ECL with no associated charged tracks within an angular cone of 15$^\circ$ measured from the IP. For the $K^0_L$ selection, we use the same criteria as in our previous study \cite{etap_belle}. We categorize $K_L^0$ candidates into three types based on clusters in the KLM and/or ECL, with different selection criteria for each of them. A cluster found only in the KLM (KLM candidate) is required to have hits in three or more KLM layers. A KLM cluster that is associated with an ECL cluster with an energy exceeding $160~\rm{MeV}$ is categorized as KLM+ECL candidate. A cluster that is found only in the ECL (ECL candidate) must have an energy above $200~\rm{MeV}$. Fake $K_L^0$'s among the KLM+ECL and ECL candidates are suppressed using additional information from the ECL: the distance between the cluster and the nearest charged track incident point on the ECL, the cluster energy, mass, width and compactness. From these, we calculate a likelihood using probability density functions (PDFs) determined from simulated events and impose restrictions on its value.

Candidates for $\eta'$ are reconstructed from the $\eta'\to\eta(2\gamma)\pi^+\pi^-$ and $\eta'\to\eta(3\pi)\pi^+\pi^-$ decay chains ($\eta'\to\rho^0\gamma$ not being used due to its low signal-to-background ratio). For the reconstructed $\eta$ and $\eta'$ invariant masses, we require $0.50 \gmm < M_\eta < 0.57 \gmm$ ($1.6\sigma$) and $0.94 \gmm < M_{\eta'} <  0.97 \gmm$ ($2.5\sigma$), respectively.

Since the energy of the $K^0_L$ cannot be measured, $M_{\rm bc}$ and $\Delta E$ cannot be calculated in the same way as for the $B^0 \to \eta' K_S^0$ candidates. 
Instead, we identify reconstructed $B^0\to\eta'K^0_L$ candidates using the momentum of the $B$ candidate in the cms, $p^*_B$, which we calculate using the reconstructed $\eta'$ momentum and the $K^0_L$ flight direction, assuming $\Delta E=0$.  

\subsection{Continuum background rejection}

For all reconstructed decay modes, the dominant background component is due to continuum events, \textit{i.e.,} $e^+e^-\to q\bar{q}$, where $q=u,d,s$ or $c$. To distinguish between $B\bar{B}$ and continuum events we use a discriminant based on the event shape analysis. Since the topology of continuum events tends to be jet-like, in contrast to the spherical $B\bar{B}$ events, we combine a set of variables that characterizes the event topology into a signal (background) likelihood variable $\mathcal{L}_{\rm sig}$ ($\mathcal{L}_{\rm bkg}$) and form the likelihood ratio $\mathcal{R}_{\rm s/b}=\mathcal{L}_{\rm sig}/(\mathcal{L}_{\rm sig}+\mathcal{L}_{\rm bkg})$. The likelihood $\mathcal{L}_{\rm sig~(bkg)}$ includes a Fisher discriminant $\mathcal{F}$ \cite{fisher} constructed from the transverse sphericity $S_\perp$ \cite{vertex}, the angle in the cms between the thrust axis\footnote{i.e., axis $\hat{n}$ that maximizes $\sum_i |\hat{p}_i\cdot \hat{n}|$, where the sum is over all considered particles momenta $\hat{p}_i$.} of the $B$ candidate and the other particles, and a set of modified Fox-Wolfram moments \cite{foxwf}. For the optimization of the separation power of the discriminant $\mathcal{F}$, a large sample of MC signal and continuum events is used. In addition, the likelihood $\mathcal{L}_{\rm sig~(bkg)}$ includes the polar angle of the $B$ meson candidate's flight direction in the cms, $\theta_B$, and, for the $\eta'\to\rho^0\gamma$ mode, the angle between the $\eta'$ meson momentum and the $\pi^+$ momentum in the $\rho^0$ meson rest frame, $\theta_H$. Both angles follow a $(1-\cos^2 \theta_{B,H})$ distribution for signal events and a flat distribution for continuum events. 

We impose loose pre-selection criteria of $\mathcal{R}_{\rm s/b}>0.5$ and $\mathcal{R}_{\rm s/b}>0.1$ for the $K^0_L$ modes and for the $[\eta'\to\rho^0\gamma,K^0_S]$ mode, respectively. In the section \ref{subsec:sig_yield}, we describe how the different $\mathcal{R}_{\rm s/b}$ distributions of signal and background events help us extract the signal yields.


\subsection{Vertex reconstruction} 
\label{sec:vertex}
Since the $B^0$ and $\bar{B}^0$ mesons are approximately at rest in the $\Upsilon(4S)$ cms, their decay time difference ($\Delta t$) can be inferred from the displacement between $z$ position of the $B_{\rm rec}$ and $B_{\rm tag}$ decay vertices ($z_{\rm rec}$ and $z_{\rm tag}$, respectively):
\begin{equation}
\Delta t \simeq \frac{z_{\rm rec}-z_{\rm tag}}{\beta\gamma c}.
\end{equation}
In order to reconstruct the decay vertex positions, we use the same algorithm that was employed in previous Belle measurements \cite{vertex}. The vertex of the $B_{\rm rec}$ meson is reconstructed by using the charged pion tracks coming either from the $\rho^0$ or $\eta'$ decay. For these tracks, we require at least one hit in the SVD $r-\phi$ strips and at least two in the SVD $z$ strips. To improve the resolution, we use an additional constraint from the beam profile at IP in the $x-y$ plane that exploits the short flight length of the $B$ meson in this view. This constraint allows for reconstruction of the vertex even in cases when only one track has a sufficient number of associated SVD hits, which happens in about 10\% of the events. The $z$ coordinate resolution for the $B_{\rm rec}$ meson vertex ranges from $70$ to $100~\mu\rm{m}$ depending on the final state and the reconstruction efficiency is about $95 \%$. To reconstruct the decay vertex of the $B_{\rm tag}$ meson, the tracks not associated with $B_{\rm rec}$ are used. The $z$-coordinate resolution for these vertices is about $120 ~\mu\rm{m}$, and the reconstruction efficiency is $\sim 93\%$.  

To reject events with poorly reconstructed vertices, we impose cuts on the reconstructed vertex quality. For vertices reconstructed with a single track, we require $\sigma_z<500~\mu\rm{m}$,  where $\sigma_z$ is the estimated error of the vertex $z$ coordinate; for multi-track vertices, we require $\sigma_z < 200~\mu\rm{m}$ and $h<50$, where $h$ is the value of the $\chi^2$ in three-dimensional space calculated using the charged tracks and without the constraint derived from the interaction region profile. In our previous analysis \cite{etap_belle}, the value of the $\chi^2$ of the vertex was only calculated along the $z$ direction, but a detailed MC study indicates that $h$ is a superior indicator of the vertex goodness-of-fit because it is less sensitive to the specific $B$ decay mode \cite{jpsi_2012}. Among events for which both the $B_{\rm rec}$ and $B_{\rm tag}$ meson vertices are reconstructed, those with $|\Delta t| < 70~\rm{ps}$ are retained for further analysis.

\subsection{Flavor tagging} 
     
The $b$-flavor of the $B_{\rm tag}$ meson is determined from inclusive properties of the particles in the event that are not associated with the $B_{\rm rec}$ reconstruction. The algorithm used is described in detail in Ref. \cite{tagging}. Two parameters, $q$ and $r$, are used to represent the tagging information. The former is the identified flavor of $B_{\rm tag}$, and can take the values $+1$ ($B^0$ tag) or $-1$ ($\bar{B}^0$ tag). The latter is an event-by-event MC-determined flavor tagging quality factor ranging from $r=0$ for no flavor discrimination to $r=1$ for unambiguously determined flavor. The data are sorted into seven intervals of $r$, and the fraction of wrongly tagged $B$ candidates in each interval, $w_l~(l=1,...,7)$, and their differences between $B^0$ and $\bar{B}^0$, $\Delta w_l$, are determined from self-tagged semileptonic and hadronic $b\to c$ decays. The total effective tagging efficiency, $\sum(f_l\times(1-2w_l)^2)$, where $f_l$ is the fraction of events in category $l$, is determined to be $0.298\pm 0.004$.

The expected $\Delta t$ distribution of signal events as given by equation (\ref{eq:sig_pdf}) is modified due to the wrong tag fraction $w$ and the wrong tag fraction difference $\Delta w$ to:
\begin{equation}
\mathcal{P}_{\rm sig}(\Delta t,q) = \frac{e^{-|\Delta t|/\tau_{B^0}}}{4\tau_{B^0}} \left( 1 -q\Delta w + (1-2w)q \cdot \left[ \Scp\sin (\Delta m_d \Delta t) + \Acp \cos (\Delta m_d \Delta t)  \right] \right).
\label{eq:sig_pdf_full}
\end{equation}

\subsection{Signal yield extraction}  
\label{subsec:sig_yield}

To determine the signal yield, we perform a multi-dimensional, unbinned, extended maximum-likelihood fit to the candidate distributions in the $M_{\rm bc}-\Delta E-\mathcal{R}_{\rm s/b}$ space for the $K^0_S$ modes, and in the $p^{*}_B-\mathcal{R}_{\rm s/b}$ space for the $K^0_L$ modes. 

\begin{figure}[tbp]
\centering 
\includegraphics[width=.325\textwidth]{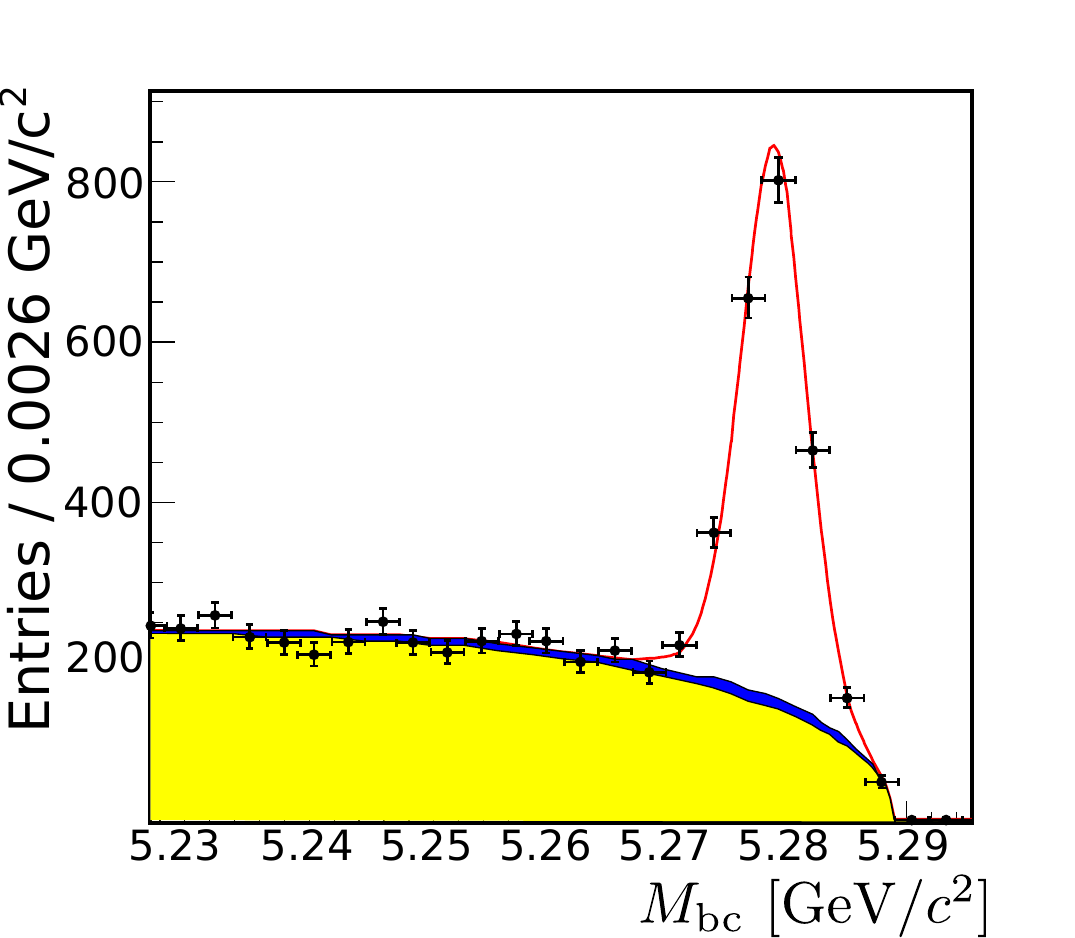}
\includegraphics[width=.325\textwidth]{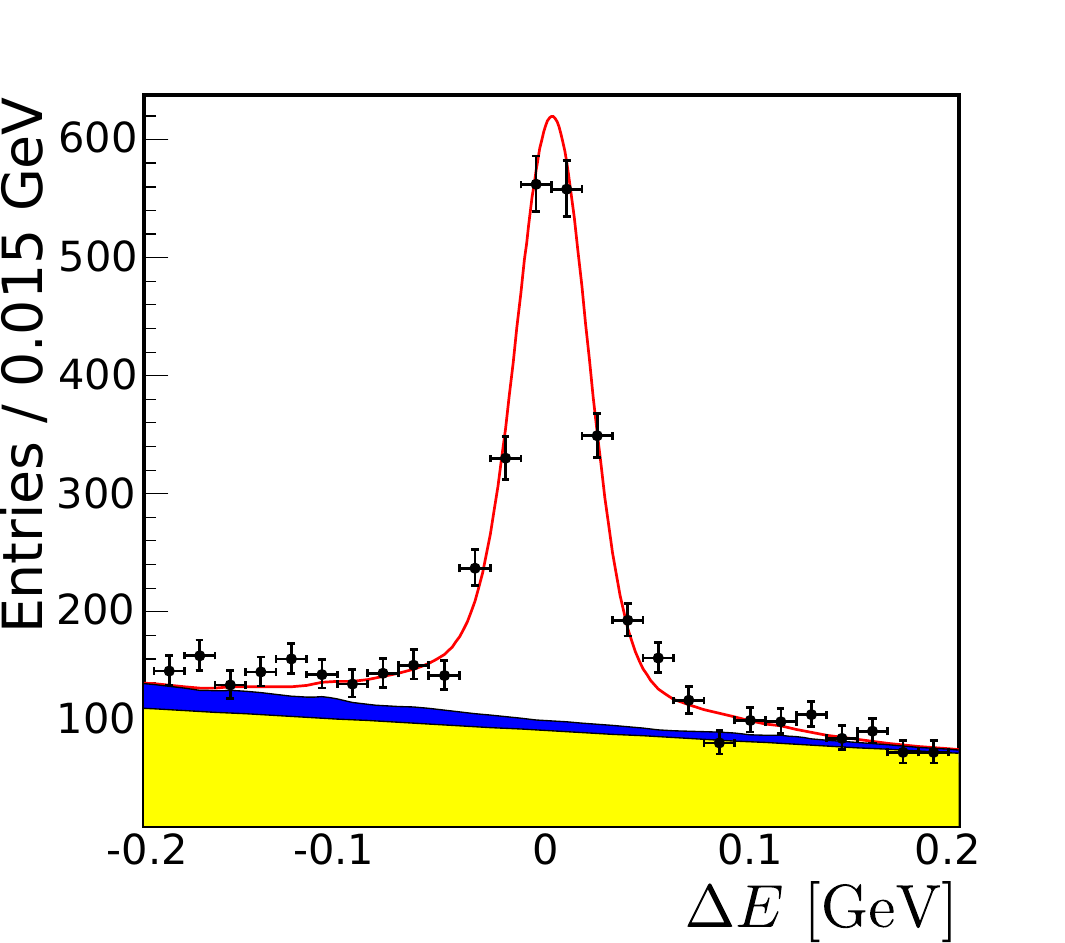}
\includegraphics[width=.325\textwidth]{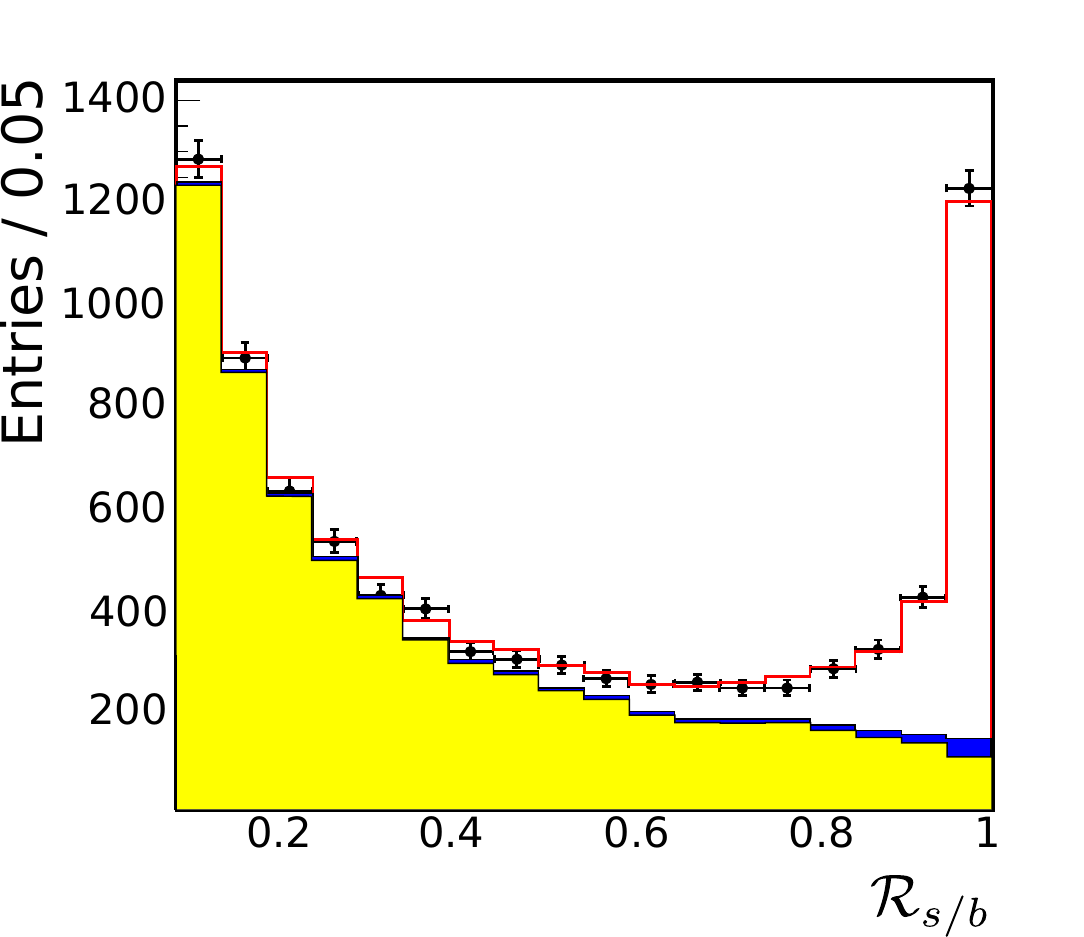}
\caption{$M_{\rm bc},~\Delta E$ and $\mathcal{R}_{\rm s/b}$ distributions of $B\to\eta'K^0_S$ candidates for all decay modes combined: the $M_{\rm bc}$ distribution for the candidates within the $\Delta E$ signal region and with $\mathcal{R}_{\rm s/b}>0.7$, the $\Delta E$ distribution for the candidates within the $M_{\rm bc}$ signal region and with $\mathcal{R}_{\rm s/b}>0.7$, and the $\mathcal{R}_{\rm s/b}$ distribution is shown for the candidates within the $M_{\rm bc}-\Delta E$  signal region. (See table \ref{tab:region} for signal region definition.) The dots with error bars show the data distribution, the red solid lines show the corresponding one-dimensional projections of the fitted model (signal+background), and the yellow and blue areas show the contributions from the continuum and $B\bar{B}$ background, respectively.}
\label{fig:frac_ks}
\end{figure}

For the $K^0_S$ modes, the fit model consists of contributions from the signal, background from continuum events, and background from $B\bar{B}$ events. We model the signal distribution in $M_{\rm bc}$ by a single Gaussian function, in $\Delta E$ by the sum of two Gaussian functions (core and outlier) and a bifurcated Gaussian function (tails), and in $\mathcal{R}_{\rm s/b}$ by a histogram PDF determined from a MC simulation of signal events. The parameters of the signal PDF are fixed at the values determined from a fit to simulated events, except for the width and mean of the $\Delta E$ core Gaussian that are kept as free parameters in the fit of the signal yield to account for any difference between the simulated and measured data. To model the distributions of the continuum background, an ARGUS function \cite{argus} is used for $M_{\rm bc}$, a linear function for $\Delta E$, and a histogram PDF for $\mathcal{R}_{\rm s/b}$. The histogram PDF is obtained from the distribution of candidates in the $M_{\rm bc}-\Delta E$ region that contains only a small fraction of signal and background $B\bar{B}$ events ($M_{\rm bc}<5.25 \gmm,~ -0.05 \mbox{ GeV} <\Delta E < 0.25 \mbox{ GeV}$). We observe a non-negligible correlation between the slope of the continuum background distribution for $\Delta E$ and the value of $\mathcal{R}_{\rm s/b}$ that we model by parameterizing the slope as a linear function of $\mathcal{R}_{\rm s/b}$. The continuum-model shape parameters are determined from the fit. The shape and the fraction (relative to signal) of background from $B\bar{B}$ events are determined from large simulated samples. A histogram PDF is used to model the distribution of this background in all three variables. The contribution from these events is small ($\sim 1\%$ of all $B$ candidates). We use the same parameterization model for all decay modes, but the model parameters are determined separately for each mode. The fits to determine the signal yield of each individual mode are performed in each of the seven $r$ (tagging-quality) regions simultaneously, using a common signal and background PDF shape. The fit range is $M_{\rm bc} > 5.22 ~\gmm$, $-0.25 \mbox{ GeV} < \Delta E < 0.25 \mbox{ GeV}$ and $\mathcal{R}_{\rm s/b}>0$. In figure \ref{fig:frac_ks}, we show signal-enhanced distributions for the $B$ candidates in $M_{\rm bc},\Delta E$ and $\mathcal{R}_{\rm s/b}$ with the corresponding one-dimensional projections of the fit superimposed.  

\begin{figure}[tbp]
\centering 
\includegraphics[width=.66\textwidth]{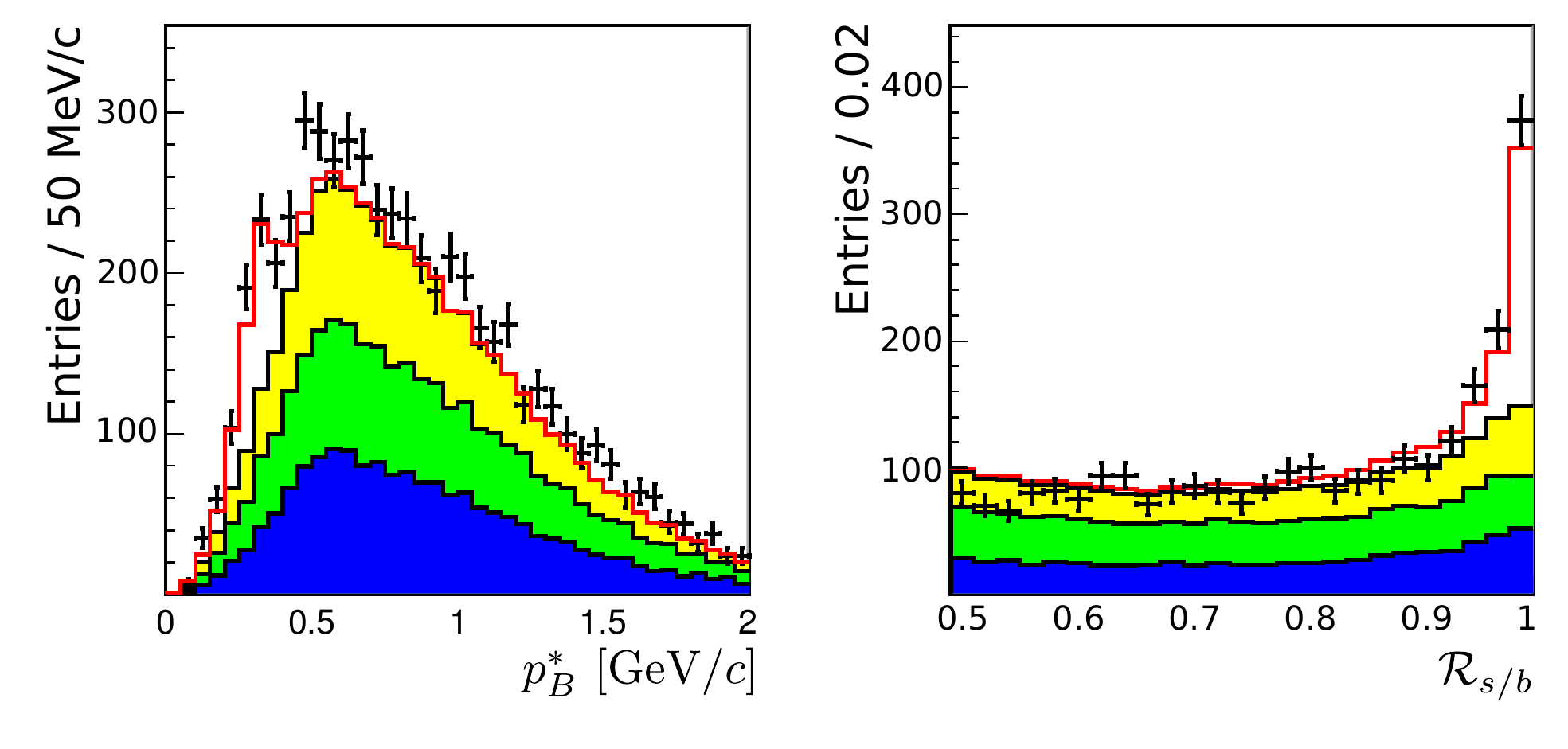}
\caption{$p^{*}_B$ and $\mathcal{R}_{\rm s/b}$ distributions of $B\to\eta'K^0_L$ candidates for all decay modes combined.  The $p^{*}_B$ ($\mathcal{R}_{\rm s/b}$) distribution is shown for the candidates within the $\mathcal{R}_{\rm s/b}$ ($p^{*}_B$) signal region. (See table \ref{tab:region} for signal region definition.) The red solid lines show the corresponding one-dimensional projections of the fitted model (signal+background). The dots with error bars show the data distribution. The blue, green, and yellow areas show the contributions from the considered background categories: fake $\eta'$, real $\eta'$ with fake $K^0_L$, and real $\eta'$ with real $K^0_L$, respectively.}
\label{fig:frac_kl}
\end{figure}

For the $K^0_L$ modes, the fit model includes the contributions of the signal and three categories of background: those with a real (correctly reconstructed) $\eta'$ and a real $K^0_L$, those with a real $\eta'$ and fake $K^0_L$ (which arise mainly from misidentification of electromagnetic showers and background neutron hits), and those with a fake $\eta'$ (combinatorial background). To model each of these contributions other than those from the fake $\eta'$, we use a histogram PDF determined from the corresponding MC sample (a signal MC for the signal and a full background sample, containing $B\bar{B}$ and continuum events, for the backgrounds). For the fake $\eta'$ contribution, the PDF and yield are estimated using reconstructed $B$ candidates with the $\eta'$ mass in the sideband regions ($0.92 \gmm < M_{\eta'} < 0.93 \gmm$, $0.97 \gmm < M_{\eta'} < 1.00 \gmm$). The PDF shape is determined and the fit is performed separately for each $\eta'$ decay mode, $K^0_L$ candidate category, and tagging-quality interval. The fit range is $p^*_B <2 ~\rm{GeV}/c$ and $\mathcal{R}_{\rm s/b} >0.5$. The reconstructed $B$-candidate distributions for $p^{*}_B$ and $\mathcal{R}_{\rm s/b}$, with the projections of the fit, are shown in figure \ref{fig:frac_kl}.

In table \ref{tab:sig_yield}, we summarize the fit-determined signal yields and sample purities in the signal region of each reconstructed decay mode. Signal region definitions are given in table \ref{tab:region}. For $K^0_S$ modes, we also show the sample purity in the region defined by $\mathcal{R}_{\rm s/b} >0.70$. This region contains about $80\%$ of the signal candidates. In total, there are $3541\pm 90$ signal candidates, where the uncertainty is statistical only. Compared to our previous measurement \cite{etap_belle}, the efficiency for reconstructing and selecting a signal event is about 15\% higher due to an increase in the track reconstruction efficiency (data reprocess) and additional 15\% is gained by a new $K^0_S$ selection method and a re-optimization of the event selection criteria.    

\begin{table}[tbp]
\centering
\begin{tabular}{cccccc}
\hline
\hline \T
& &  \multicolumn{2}{c}{Signal region} & & \multicolumn{1}{c}{$ + \mathcal{R}_{\rm s/b} >0.70$} \\
\cline{3-4} \cline{6-6} \T
$K^0$ mode & $\eta'$ mode & $N_{\rm sig}$ & Purity & & Purity \T \B \\
\hline \T \B
$K^{0}_S\to\pi^+\pi^-$ & $\rho^0\gamma$      & $ 1411\pm 48$  & $0.19$  & & $0.55$  \T \B  \\
 & $\eta(\gamma\gamma)\pi^+\pi^-$   & $648\pm 28 $  & $0.49$  & & $0.87$ \B \\
 & $\eta(3\pi)\pi^+\pi^-$     & $174\pm 14 $   & $0.65$  & & $0.93$ \B \\
\hline
$K^{0}_S\to\pi^0\pi^0$ & $\rho^0\gamma$ & $162\pm 21$ & $0.04$ & & $0.13$ \T \B \\
 & $\eta(\gamma\gamma)\pi^+\pi^-$ & $104\pm 14$ & $0.16$ & & $0.64$ \B \\ 
\hline
$K_L$ & $\eta(\gamma\gamma)\pi^+\pi^-$ & $829\pm 54$ & $0.30$ & \T \B \\ 
     & $\eta(3\pi)\pi^+\pi^-$ & $213 \pm 36$ & $0.19$ & \B \\
\hline
Total & & $3541 \pm 91$ & & \T\\
\hline
\hline
\end{tabular}
\caption{Measured signal yields $N_{\rm sig}$ and signal purities in the signal region for each $B^0\to\eta'K^0$ reconstructed decay mode. The uncertainties given are statistical only.}
\label{tab:sig_yield}
\end{table}
\section{Results of the {\it CP} asymmetry measurements}
\label{sec:result}
\subsection{Determination of the {\it CP} violation parameters}
We determine $\seff$ and $\Acp$ by performing an unbinned, maximum-likelihood fit to the observed $(\Delta t,q)$ distributions of the reconstructed signal region events. The PDF for the signal distribution, $\mathcal{P}(\Delta t,q; \seff,\Acp,w_l,\Delta w_l)$, is given by equation (\ref{eq:sig_pdf_full}) with the replacement of $\Scp$ with $-\xi_f\seff$. To account for the finite resolution of the $\Delta t$ measurement, this PDF is convolved with the $\Delta t$ resolution function, $R_{\rm sig}(\Delta t)$, which is itself a convolution of four components: the detector resolutions for both $z_{\rm rec}$ and $z_{\rm tag}$, the smearing of the $z_{\rm tag}$ vertex due to secondary tracks from charmed particle decays, and the smearing due to the kinematic approximation that the $B$ mesons are at rest in the cms. The shape of the resolution function is determined on an event-by-event basis by changing its parameters according to the vertex-fit quality indicators $h$ and $\sigma_z$ (described in section \ref{sec:vertex}).
 We use the same $R_{\rm sig}(\Delta t)$ for all $\eta'$ and $K^0$ decay modes. 
This procedure is validated by performing a fit to obtain the $B$ meson lifetime for each decay mode, as described in section \ref{sec:valid}. The resolution function $R_{\rm sig}(\Delta t)$ is described in more detail in Ref. \cite{res_fun}. 

\begin{table}[tbp]
\centering
\begin{tabular}{clcccc}
\hline
\hline \T 
$K^0$ mode & $\eta'$ mode &  $\Delta E$ [GeV] & $M_{\rm bc}$ [GeV/$c^2$] & $p_B^{*}$ [GeV/$c$] & $\mathcal{R}_{\rm{s/b}}$ \B \\
\hline 
$K^0_S\to\pi^+\pi^-$ & $\eta'\to\rho^0\gamma$ & $[-0.07,~ 0.07]$ & $>5.27$ & -- & >0.1 \T \B \\
 & $\eta'\to\eta(\gamma\gamma)\pi^+\pi^-$ & $[-0.10,~0.08]$ & $>5.27$ & -- & -- \B \\
&  $\eta'\to\eta(3\pi)\pi^+\pi^-$ & $[-0.08,~ 0.06]$ & $>5.27$ & -- & -- \B \\
$K^0_S\to\pi^0\pi^0$ & $\eta'\to\rho^0\gamma$ & $[-0.15,~0.10]$ &  $>5.27$ & -- & >0.1 \B \\
& $\eta'\to\eta(\gamma\gamma)\pi^+\pi^-$ & $[-0.15,~0.10]$ &  $>5.27$ & -- & -- \B \\
$K^0_L$ & all & -- & -- & $[0.20,~ 0.45]$ & >0.9\B \\
\hline
\hline
\end{tabular}
\caption{Signal region definitions for all reconstructed decay modes.}
\label{tab:region}
\end{table}

We assign the following likelihood value to each event:
\begin{align}
P_i(\Delta t_i,q_i) = (1-f_{\rm ol})\int &\Big[~f^i_{\rm sig}\mathcal{P}_{\rm sig}(\Delta t',q_i)R_{\rm sig}(\Delta t_i - \Delta t') 
  \nonumber \\ &+ \sum_k f^i_k\mathcal{P}^k_{\rm bkg}(\Delta t') R^k_{\rm bkg}(\Delta t_i -\Delta t')~\Big]~ d(\Delta t') + f_{\rm ol}P_{\rm ol}(\Delta t_i),
\end{align}
where $P_{\rm ol}(\Delta t)$ is a broad Gaussian function ($\sigma \sim 40\mbox{ ps}$) with a small fraction $f_{\rm ol}$ of $\mathcal{O}(10^{-3})$ that represents an outlier component caused by wrongly reconstructed vertices \cite{res_fun}. The sum $\sum_k$ runs over all considered background categories, as defined in the previous section. The signal and background-category probabilities, $f^i_{\rm sig}$ and $f^i_{k}$, respectively, depend on the $r$ interval and are calculated on an event-by-event basis as functions of $M_{\rm bc},~\Delta E$ and $\mathcal{R}_{\rm s/b}$ for the $K^0_S$ modes, and as functions of $p^*_B$ and $\mathcal{R}_{\rm s/b}$ for the $K^0_L$ modes. The function shapes are determined by the fit described in the previous section. A PDF for each background category, $\mathcal{P}^k_{\rm bkg}(\Delta t)$, is modeled as the sum of prompt and exponential decay components parameterized as a Dirac delta function and $e^{-|\Delta t|/\tau_{\rm bkg}}$, respectively, where $\tau_{\rm bkg}$ is the effective lifetime in background events. All background PDFs are convolved with a background resolution function, $R^k_{\rm bkg}$, that is modeled as the sum of three (two) Gaussian functions for the $K^0_S$ ($K^0_L$) modes. For the $K^0_S$ modes, the shape parameters of the continuum background PDF are determined by a fit to the $\Delta t$ distribution of events in the $M_{\rm bc} - \Delta E-\mathcal{R}_{\rm s/b}$ region containing a very small fraction ($<1$\%) of $B\bar{B}$ events ($M_{\rm bc} < 5.265 \gmm, -0.1 \mbox{ GeV} < \Delta E < 0.25 \mbox{ GeV},\mathcal{R}_{\rm s/b} < 0.9$), and for the $B\bar{B}$ background PDF by the fit to the $\Delta t$ distribution of events from the MC simulation. For the $K^0_L$ modes, the background PDF shape parameters are determined by a fit to off-resonance data for the background arising from continuum events; a fit to events in the $\eta'$ mass sideband is used for the fake $\eta'$ background category. Relatively small background contributions ($\sim 1\%$) from $B\to\eta'X$ decays 
are included in the background PDF. Their fractions and $\Delta t$ PDF parameters are obtained from large corresponding MC samples. 

\begin{figure}[tbp]
\centering 
\includegraphics[width=.45\textwidth]{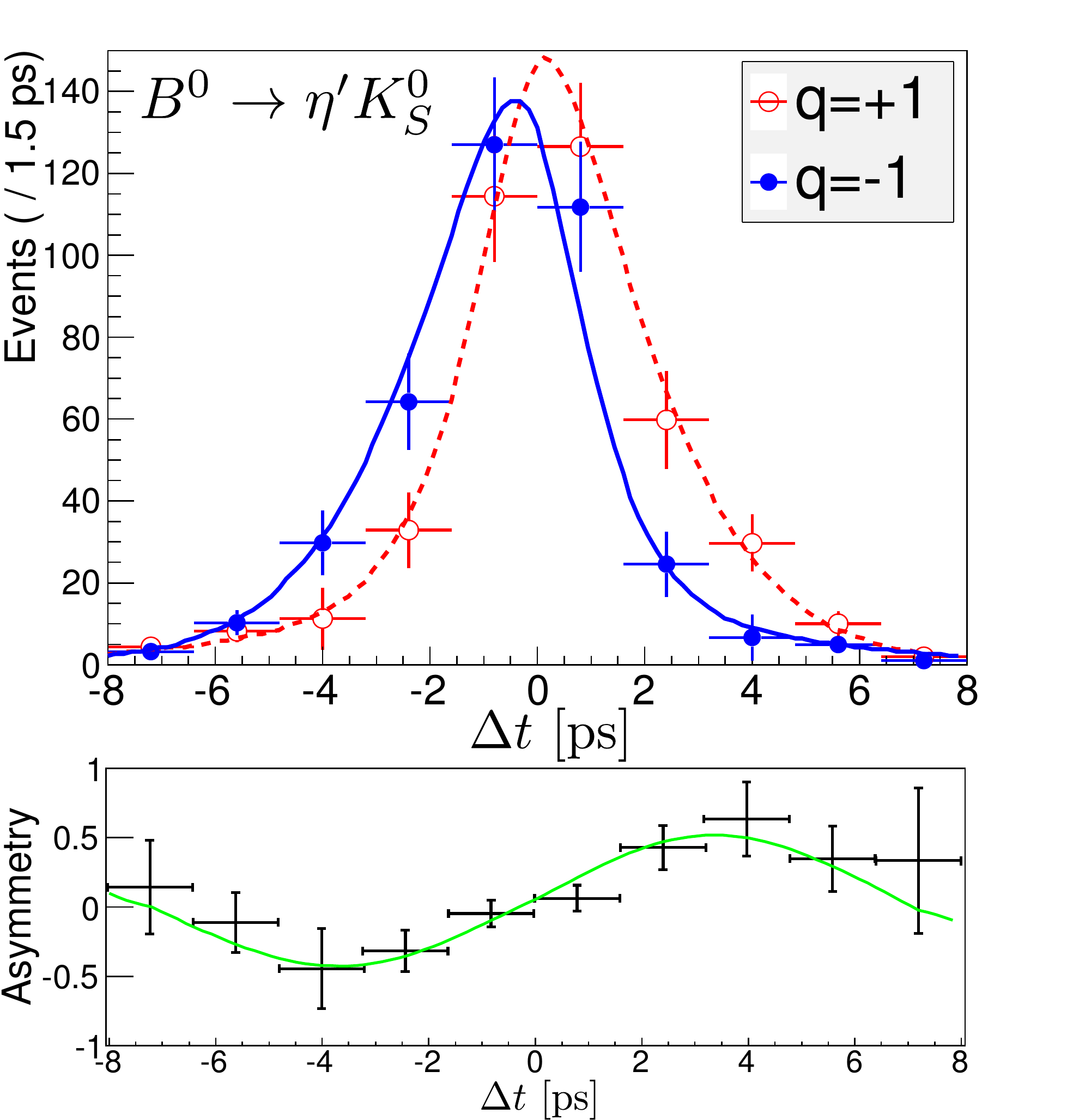}
\includegraphics[width=.46\textwidth]{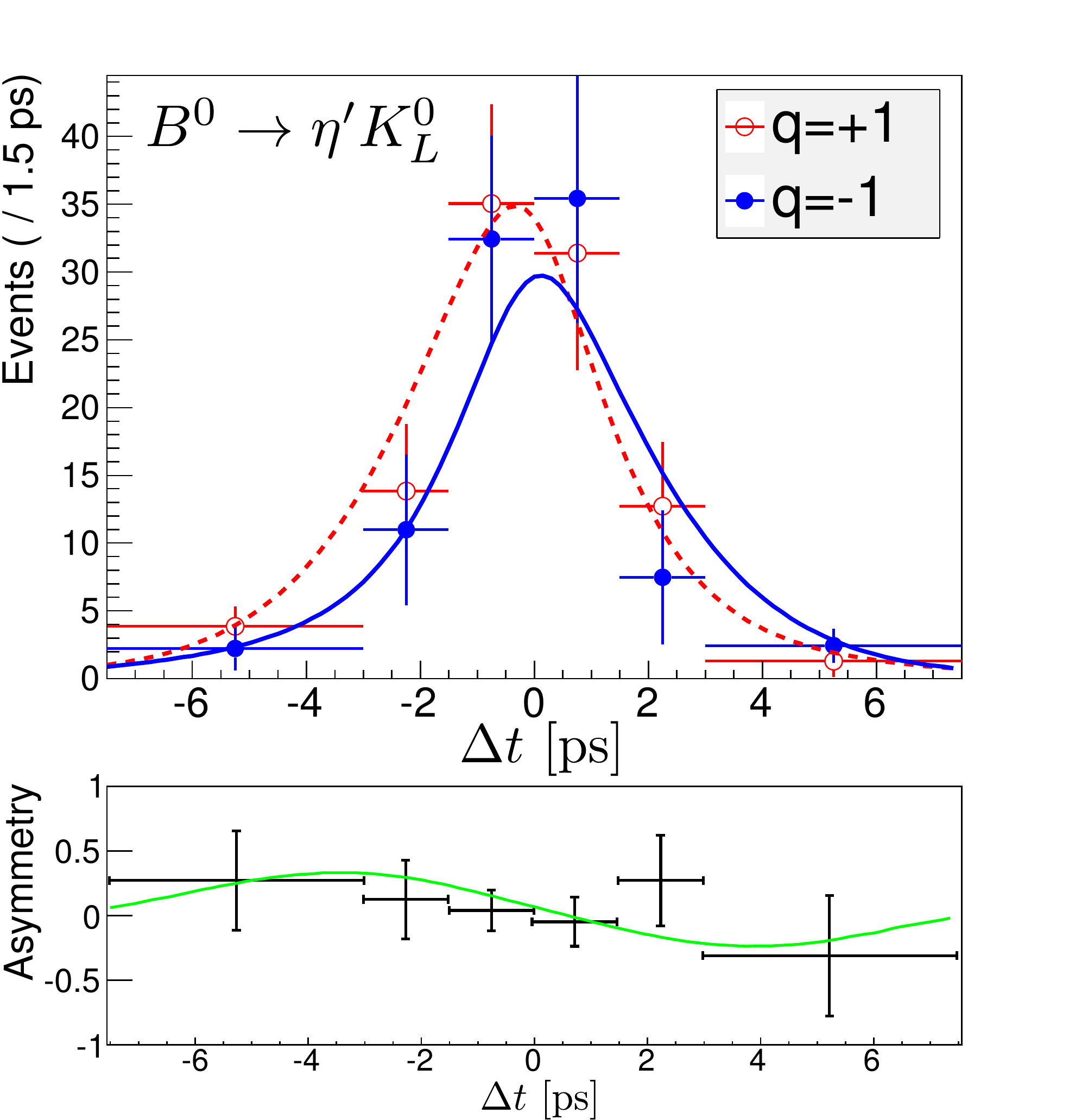}
\caption{Background-subtracted $\Delta t$ distribution (top) and asymmetry (bottom) for $B^0\to\eta'K^0_S$ (left) and $B^0\to\eta'K^0_L$ (right) events with good flavor tags ($r>0.5$). All reconstructed modes are combined. The solid curves show the fitted PDF, and the points with error bars show the data distributions.}
\label{fig:asym}
\end{figure}

In the fit, we fix $\tau_{B^0}$ and $\Delta m_d$ to their current world-average values of 1.519 ps and $0.51\times 10^{12}~ \hbar\mbox{s}^{-1}$, respectively \cite{pdg}. The only free parameters in the final fit are $\seff$ and $\Acp$, which are determined by maximizing the likelihood function $L=\prod_i P_i(\Delta t,q;\seff,\Acp)$, where the product spans all candidate events. We maximize this likelihood for the $K^0_S$ and $K^0_L$ decay modes individually, as well as simultaneously, taking into account their different {\it CP}-eigenstate values. The fit results are shown in table \ref{tab:fit_result}. We define the background-subtracted asymmetry in each $\Delta t$ bin by $(N_+ - N_-)/(N_+ + N_-)$, where $N_+ ~(N_-)$ is the signal yield with $q=+1~(-1)$. This asymmetry and the background-subtracted $\Delta t$ distribution are shown in figure \ref{fig:asym}. 
\begin{table}[tbp]
\centering
\begin{tabular}{ccc}
\hline
\hline \T 
Decay mode  &  $\seff$ & $\mathcal{A}_f$ \B \\
\hline 
$\eta' K^0_S$ & $+0.71 \pm 0.07$ & $+0.02\pm 0.05$ \T \B \\
$\eta' K^0_L$ & $+0.46\pm 0.21$  & $+0.09\pm 0.14$ \B \\
$\eta'K^0$ & $+0.68\pm 0.07 \pm 0.03$ & $+0.03\pm 0.05\pm 0.04$ \B \\ 
\hline
\hline
\end{tabular}
\caption{Results of the fits to the $(\Delta t,q)$ distributions. For the separate fits of $\eta'K^0_S$ and $\eta'K^0_L$ sub-samples, only statistical uncertainties are given; for the combined fit the first uncertainty is statistical and the second is systematic (see section \ref{sec:sys}).}
\label{tab:fit_result}
\end{table}

\subsection{Systematic uncertainty}
\label{sec:sys}
The systematic uncertainties are listed in table \ref{tab:syst}, where the total systematic uncertainty is obtained by adding all contributions in quadrature. Uncertainties originating from the vertex reconstruction algorithm are a significant part of the systematic uncertainty for both $\seff$ and $\Acp$. They are estimated by varying several algorithm conditions and repeating the final fit. The variation of the results with respect to the nominal result is taken as a systematic uncertainty. In particular, the effect of the vertex quality criteria is estimated by changing the requirements $h<50$ to either $h<25$ or $h<100$ and $\sigma_{z}<200~(500)~\mu\rm{m}$ to $\sigma_{z}<150~(400)~\mu\rm{m}$ or $\sigma_{z}<300~(600)~\mu\rm{m}$ for multi-track (single-track) vertices. The systematic uncertainty due to the IP constraint in the vertex reconstruction is estimated by varying the IP profile size in the plane perpendicular to the $z$-axis. The effect of the criteria for the selection of tracks used in the $B_{\rm tag}$ vertex reconstruction is estimated by changing the requirement on the distance of closest approach with respect to the reconstructed vertex by $\pm 100~\mu\rm{m}$ from the nominal maximum value of 500 $\mu\rm{m}$. Small biases in the $\Delta z$ measurement are observed in $e^+e^-\to\mu^+\mu^-$ and other control samples; to account for these, a special correction function is applied and the fit is repeated. To estimate the uncertainty due to the $|\Delta t|$ fit range, we vary the requirement $|\Delta t|<70\mbox{ ps}$ by $\pm 30 \mbox{ ps}$. For the systematic uncertainties due to an imperfect SVD alignment, we use the value from the latest Belle $\sin2\phi_1$ measurement \cite{jpsi_2012}, estimated from MC samples with artificial misalignment effects. The largest contribution to the $\seff$ uncertainty arises from the uncertainties in the $\Delta t$ resolution function parameters. We vary each of the parameters obtained from data (MC) by $\pm 1\sigma$ ($\pm 2\sigma$)\footnote{In order to include possible systematic differences between the measured and simulated data we vary the parameters obtained from the latter for $\pm 2\sigma$.}, repeat the fit, and add the variations in quadrature. Similarly, we estimate the contributions from the uncertainties in the extracted signal fractions, the background $\Delta t$ distributions, and physics parameters $\tau_B^0$ and $\Delta m_d$. Systematic errors due to uncertainties in the wrong tag fractions are studied by varying the wrong tag fraction individually in each $r$ region. A possible fit bias is examined by fitting a large number of MC events. Finally, we estimate the contribution from the effect of the tag-side interference \cite{tsi}, which introduces a significant contribution to the systematic uncertainty of $\Acp$. Interference between the CKM-favored and -suppressed $B\to D$ transitions in the $f_{\rm tag}$ final state results in a small correction to the PDF for the signal $\Delta t$ distribution. The size of the correction is estimated using the $B^0\to D^{*-}l^+\nu$ sample; then, a set of MC pseudo-experiments is generated and an ensemble test is performed to obtain possible systematic biases in $\seff$ and $\Acp$.

\begin{table}[tbp]
\centering
\begin{tabular}{lcc}
\hline
\hline \T
Source  &  $\mathcal{S}_{\eta'K^0}$ & $\mathcal{A}_{\eta'K^0}$ \B \\
\hline 
Vertexing & $\pm 0.014$   & $\pm 0.033$   \T\B \\
$\Delta t$ resolution & $\pm 0.025$ & $\pm 0.006$ \B\\
$\eta'K^0_S$ signal fraction & $\pm 0.013$ & $\pm 0.006$ \B \\
$\eta'K^0_L$ signal fraction & $\pm 0.005$ & $\pm 0.004$ \B \\
Background $\Delta t$ PDF\hspace{0.5cm} & $\pm 0.001$ & $< 0.001$ \B \\
Physics parameters & $\pm 0.001$ & $< 0.001$ \B \\
Flavor tagging & $\pm 0.003$ & $\pm 0.003$ \B \\
Possible fit bias & $\pm 0.001$ & $ \pm 0.001$ \B \\
Tag-side interference & $\pm 0.001$ & $\pm 0.020$ \B \\
\hline \T
Total &  $\pm 0.032$   &   $\pm 0.040$ \T \B \\
\hline
\hline
\end{tabular}
\caption{Summary of systematic uncertainties affecting $\seff$ and $\Acp$.}
\label{tab:syst}
\end{table}
\subsection{Validation tests}
\label{sec:valid}
Various cross-checks of the measurement are performed. We fit a large number of independent MC samples containing the expected number of signal and background events. We observe no significant bias between the generated and fitted values of the $CP$ violation parameters and confirm the linear response of the fitter. The measurement method is tested by using measured data. Since only charged tracks from $\eta'$ are used for the decay vertex reconstruction, we reconstruct charged $B$ meson decays, $B^+\to\eta'K^+$, for which the obtained signal yield is about three times larger than for $B^0\to\eta'K^0$. We perform lifetime measurements with both $B^0\to\eta'K^0$ and $B^+\to\eta'K^+$ data samples, using the same procedure as for the measurement of the $CP$ violation parameters apart from the flavor tagging, which is not used. The fit yields $\tau_{B^0}$ and $\tau_{B^+}$ values consistent with the world-average values (we measure $\tau_{B^+}=1.65\pm 0.03$ ps and $\tau_{B^0} = 1.49 \pm 0.04$ ps, while the world-average values are $\tau_{B^+}=1.641\pm 0.008$ ps and $\tau_{B^0} = 1.519 \pm 0.007$ ps \cite{pdg}). We also apply our nominal fit procedure to the charged data sample. The results obtained for $\mathcal{S}_{\eta'K^+}$ and $\mathcal{A}_{\eta'K^+}$ are consistent with no ${\it CP}$ asymmetry, as expected (we measure $\mathcal{S}_{\eta'K^+} = -0.04\pm 0.03$ and $\mathcal{A}_{\eta'K^+}=0.00\pm 0.02$, where the uncertainties are statistical only). Finally, by the use of MC pseudo-experiments, we confirm that the statistical uncertainties obtained in our measurement are consistent with the expectations from the ensemble test. 

\section{Summary}
\label{sec:summary}

Using the full Belle $\Upsilon(4S)$ data set containing $772\times 10^6~B\bar{B}$ pairs, a measurement of the time-dependent {\it CP} violation parameters in $B^0\to\eta'K^0$ decays is performed. We fit the $(\Delta t,q)$ distributions of the reconstructed $B^0\to\eta'K^0_S$ and $B^0\to\eta'K^0_L$ candidates, and obtain
\begin{align}
\seff &= +0.68\pm 0.07 \pm 0.03,\nonumber\\
\Acp &= +0.03\pm 0.05\pm 0.04,\nonumber
\end{align}
where the first uncertainties are statistical and the second systematic. These results are consistent with and supersede our previous measurement \cite{etap_belle}. They are the most precise determination of these parameters to date and are consistent with the world-average value of $\sin 2\phi_1$ obtained from the $B^0\to J/\psi K^0$ decay \cite{pdg}. No deviations from the predictions of the Standard Model are observed.






\acknowledgments

We thank the KEKB group for the excellent operation of the
accelerator; the KEK cryogenics group for the efficient
operation of the solenoid; and the KEK computer group,
the National Institute of Informatics, and the 
PNNL/EMSL computing group for valuable computing
and SINET4 network support.  We acknowledge support from
the Ministry of Education, Culture, Sports, Science, and
Technology (MEXT) of Japan, the Japan Society for the 
Promotion of Science (JSPS), and the Tau-Lepton Physics 
Research Center of Nagoya University; 
the Australian Research Council and the Australian 
Department of Industry, Innovation, Science and Research;
Austrian Science Fund under Grant No. P 22742-N16;
the National Natural Science Foundation of China under Contracts 
No.~10575109, No.~10775142, No.~10825524, No.~10875115, No.~10935008 
and No.~11175187; 
the Ministry of Education, Youth and Sports of the Czech
Republic under Contract No.~LG14034;
the Carl Zeiss Foundation, the Deutsche Forschungsgemeinschaft
and the VolkswagenStiftung;
the Department of Science and Technology of India; 
the Istituto Nazionale di Fisica Nucleare of Italy; 
the WCU program of the Ministry of Education, Science and
Technology, National Research Foundation of Korea Grants
No.~2011-0029457, No.~2012-0008143, No.~2012R1A1A2008330,
No.~2013R1A1A3007772;
the BRL program under NRF Grant No.~KRF-2011-0020333,
No.~KRF-2011-0021196,
Center for Korean J-PARC Users, No.~NRF-2013K1A3A7A06056592; the BK21
Plus program and the GSDC of the Korea Institute of Science and
Technology Information;
the Polish Ministry of Science and Higher Education and 
the National Science Center;
the Ministry of Education and Science of the Russian
Federation and the Russian Federal Agency for Atomic Energy;
the Slovenian Research Agency;
the Basque Foundation for Science (IKERBASQUE) and the UPV/EHU under 
program UFI 11/55;
the Swiss National Science Foundation; the National Science Council
and the Ministry of Education of Taiwan; and the U.S.\
Department of Energy and the National Science Foundation.
This work is supported by a Grant-in-Aid from MEXT for 
Science Research in a Priority Area (``New Development of 
Flavor Physics'') and from JSPS for Creative Scientific 
Research (``Evolution of Tau-lepton Physics'').



\end{document}